\documentclass[11pt,twoside]{article}
\usepackage{graphicx,epsfig,natbib}
\usepackage[suffix=]{epstopdf}
\usepackage{CS18}

\begin{document}
%
%

\title{Using Transiting Planets to Model Starspot Evolution}

\author{James R. A. Davenport$^{1}$, Leslie Hebb$^{2}$, Suzanne L. Hawley$^{1}$}

\affil{$^1$Department of Astronomy, University of Washington, Box 351580, Seattle, WA 98195}
\affil{$^2$Department of Physics, Hobart and William Smith Colleges, Geneva, NY, 14456}

\begin{abstract}
Photometry from {\it Kepler}\index{Kepler} has revealed the presence of cool starspots\index{starspots} on the surfaces of thousands of stars, presenting a wide range of spot morphologies and lifetimes. Understanding the lifetime and evolution of starspots across the main sequence reveals critical information about the strength and nature of stellar dynamos. We probe the dynamo by modeling the starspot properties over time using {\it Kepler} light curves. In particular, we use planetary systems like Kepler 17 that show in-transit starspot crossing features. 
Spot-occulting transits probe smaller-scale starspot features on the stellar surface along a fixed latitude region. Our approach is novel in modeling both the in- and out-of transit light curve features, allowing us to break fundamental degeneracies between spot size, latitude, and contrast. With continuous monitoring from {\it Kepler} we are able to observe small changes in the positions and sizes of spots from many transits, spanning 4 years of data. Additionally, for stars without transiting planets like GJ 1243, we are able to recover subtle, long term changes in spot sizes and longitudes, leading to some of the slowest differential rotation\index{differential rotation} rates yet measured. These studies constrain key physical parameters including rotation period, differential rotation, and diffusion timescales, and open the door to ensemble studies of detailed spot evolution in the future.
\end{abstract}

\section{Introduction: Modeling Starspots in Kepler}

With the recent availability of extremely high precision light curves from {\it Kepler} and CoRoT of many transiting planet host stars \citep{borucki2010,barge2008}, we can now observe local magnetic field features on a significant number of stars other than the Sun by detecting and characterizing individual starspots and spot groups. Since Sunspots are known to be the location of strong concentrations of magnetic fields on the Sun, individual starspots can be used as tracers of the underlying small-scale magnetic fields on low mass stars.

A method for characterizing small spotted regions by identifying and modeling brightness variations during planetary transits was first described in \citet{silva2003}. An apparent increase in brightness that appears as a ``bump'' in the light curve during a planetary transit is the signature of a planet crossing in front of a small starspot that is cooler and darker than the surrounding stellar surface.  Such anomalous brightness variations are detectable in high precision photometry with sub-millimag precision, and have now been observed on several known planet host stars (e.g. CoRot-2, Tres-1, HAT-P-11, GJ 1214, Kepler 17). The planet acts as a ``knife-edge'' probe of the stellar surface, highlighting brightness variations on the star and providing precise positional information about the spots in the path of the planet. Since the transits are periodically repeating and {\it Kepler} provides continuous photometric coverage, 
the occulted spots are detected multiple times, and it is possible to observe their emergence and decay.

Starspot lifetime is related to the magnetic flux diffusion timescale.  This parameter has been measured to high precision in the Sun \citep[e.g.][]{schrijver2001}, but no robust measurements currently exist for other stars. Thus, the majority of stellar dynamo and magnetic flux transport models adopt Solar values, even when attempting to model stars with fundamentally different properties (i.e. mass, dynamo type, convection zone depth). By making measurements of starspot lifetime for stars with a range of masses and rotation rates, we constrain this important property as a function of convection zone depth and empirically determine the effect of rotation on the flux diffusion.

We recently completed the development of an eclipse mapping program designed to derive the
starspot distribution on the surface of a transiting planet host star which shows in-transit starspot crossing features in its light curve.  In general, light curve modeling is typically very sensitive to
the longitudes of active regions, but information about the latitudes of the spots is difficult to recover.
Furthermore, there are many degenerate solutions when converting a 1D integrated brightness light curve into a 2D surface brightness distribution. However, our transiting systems are
unique in that they provide additional information about specific spot latitudes and longitudes
in the regions that are occulted by the planets.  This reduces the degeneracy of the overall problem
and allows us to measure precise positions of certain starspots.

A light curve of a transiting planet host star obtained by {\it Kepler}
is read into the program along with the physical and orbital properties of the star and planet.
The goal is to derive a set of properties (latitude, longitude and radius) for all the spots that
best reproduces the observed light curve through a $\chi^{2}$ comparison between the observed data and a synthetic light curve. In our code, we simulate the star as a uniform surface brightness sphere
with limb darkening on which circular, gray starspots are fixed.  Any number of planets may orbit the star.
The star is defined to have a radius of 1.0 and a known rotation period ($P_{rot}$).
Each planet is defined by its orbital period ($P_{orb}$), time of mid transit ($T_0$), impact parameter ($b$), radius and orbital separation relative to the size of the host star ($R_p/R_s$ and $a/R_s$), eccentricity ($ecc$) and argument of periastron ($\omega$). Each spot is characterized by its position (longitude and latitude) on the stellar surface, its radius and its average brightness (0 = totally dark, 1 = same brightness as the star).

At each time step in the input light curve, as the star rotates and the planet orbits, the code calculates the flux from the star, given the light blocked by the orbiting planets and the spots rotating in and out of view.
It also determines whether the planet is overlapping any of the spots. Limb darkening is implemented by treating the star as a series of overlapping, concentric circles with brightness values defined by the 4-coefficient limb darkening model \citep{claret2011}.
This model light curve generating engine is wrapped with an affine invariant Markov Chain Monte Carlo (MCMC) sampler based on \citet{emcee}, which explores the parameter space and derives the optimum spot properties. The program can work for a planetary system with one or more
planets of any size, and systems in which the spin-axis of the star and orbital axis
of the planet are in any orientation (aligned or misaligned).
However, the program requires that we choose the number of spots on the star a~priori.  It also needs
a well defined transit light curve model that provides the planet and orbital properties and the rotation
period of the star.  
Furthermore, the current version of our code requires the spot distribution to remain static.  Therefore, we only model a subset of the
data at any one time.  We adopt a time duration or ``window" to model over which we do not expect the
spots to evolve.  By sliding this ``window" over the full length of the light curve and running the code many times, we fit the entire light curve and determine the secular evolution of the spots.

In this contributed talk we presented examples of our code in action, modeling the starspots on two target stars (GJ 1243 and Kepler 17).

\section{Stars Without Transiting Exoplanets}

The four years of {\it Kepler} data have revealed the presence of starspots on more than 40,000 stars \citep{reinhold2013}. These spots are detected en masse by the modulation of the light curves as the cool spots rotate in and out of view. This provides a robust measure of the stellar rotation period, averaged over the entire light curve. Deeper amplitude flux modulations indicate larger starspots, producing first-order estimates for the starspot coverage on every star. However, to measure differential rotation and spot evolution we must precisely track the size and position of these spot features over time. 

For our initial study we have focused on the highly active M4 dwarf, GJ 1243. This cool star has a rapid rotation period of 0.5926 days \citep{savanov2011}, and displayed more than 6000 flares over the 11 months of {\it Kepler} 1-minute ``short-cadence'' data \citep{hawley2014,davenport2014b}. For our starspot analysis we have used the {\it Kepler} long-cadence (30-minute) data, which diminishes the impact of small amplitude flares on the smooth starspot modulations. We utilized 14 quarters of long-cadence data, spanning the entire four year {\it Kepler} mission.

As \citet{savanov2011} point out, the GJ 1243 light curve shows two clear starspot signatures. For the entire duration of the GJ 1243 light curve we have modeled the locations and sizes for two spots, using 5 day windows for each run of our model. We advanced each subsequent time window by 2.5 days, and required each window to contain at least 100 epochs to be fit with our model. For systems with no transiting planet, such as GJ 1243, fundamental degeneracies exist in constraining the latitudes and radii of the spots. We have fixed the latitudes of our two starspots, placing one along the stellar equator and the other at a latitude of $\sim$40$^\circ$.

\begin{figure}
\makebox[\textwidth][c]{\includegraphics[width=7in]{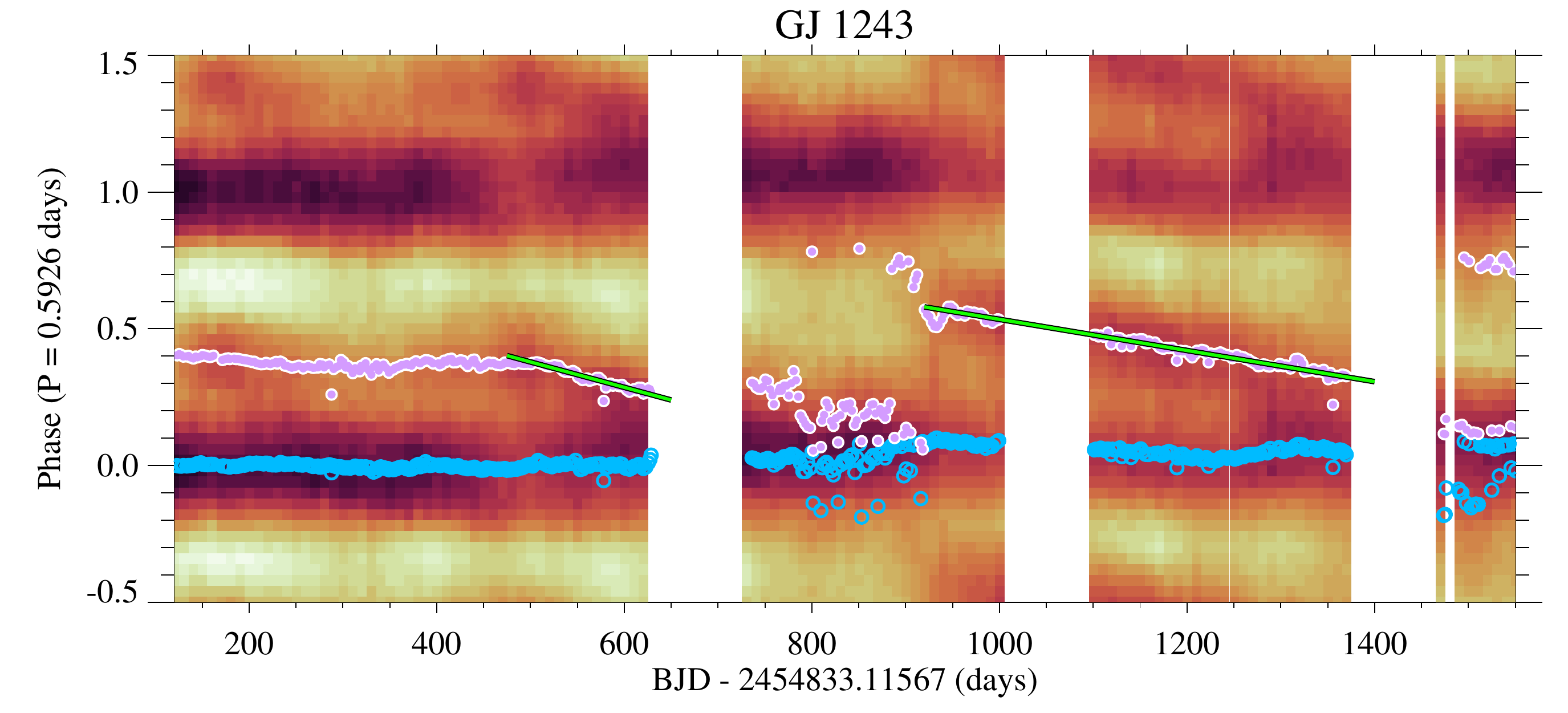}}
\caption{Continuous phased light curve map for GJ 1243, with the best-fit solutions from our starspot model run overlaid (open and filled circles). Pixel shade (dark to light) corresponds with increasing flux levels. For clarity the light curve is folded twice in phase. The higher latitude spot (open blue circles) remains nearly constant in phase, while the secondary lower latitude spot (purple filled circles) evolves significantly. Linear fits to the phase evolution for the secondary spot are overlaid (solid green lines), which we interpret as differential rotation.}
\label{gj1243}
\end{figure}

In Figure \ref{gj1243} we present the flux of GJ 1243, mapped as a function of both the rotation phase (or equivalently longitude) and time. This temporal ``phase map'' shows the starspot as dark bands, which evolve in both amplitude (shading) and phase over time. The starspots have a typical flux modulation amplitude of 2\% flux. Several starspots features are present throughout the phase map. The larger amplitude ``primary'' starspot is centered at Phase = 0, and is nearly constant in amplitude and phase throughout the four years. A smaller amplitude feature is seen, typically near Phase = 0.4, which evolves significantly in both phase and amplitude. This ``secondary'' spot nearly disappears at times BJD$\sim$300 and $\sim$800.

We have also overlaid in Figure \ref{gj1243} the best-fit longitude results from our two-starspot MCMC model. The higher-latitude spot in our model is consistently associated with the Phase = 0 primary starspot, while the equatorial spot evolves with the secondary feature. The secondary starspot at times appears to evolve linearly in longitude, which we interpret as the signature of differential rotation.  Linear fits are shown for two such phase evolutions in Figure \ref{gj1243}. This differential rotation rate is exceptionally small, leading to estimated equator-lap-pole times nearly an order of magnitude longer than seen on the Sun. This is the slowest rate of differential rotation for a main sequence star ever measured.

\clearpage

\section{Stars With Transiting Exoplanets}
While we have been able to derive a great amount of information about the starspot sizes, positions, and longitudinal evolution for stars without transits, this analysis is essentially limited to only 2 or 3 spots within a given time window. In the case of stars {\it with} a transiting planet, we search for ``bumps'' in the light curve within the transit, signifying the planet passing over region that is cooler (fainter) than the surrounding stellar surface. As described above, these bumps allow us to detect much smaller scale features on the star along the projected path of the planet's transit. As the star spins, each subsequent transit probes a different set of surface features.

In the case of Kepler 17, a Solar-like G2 dwarf, bumps can be seen clearly in many transits. The planet's orbit is alined with the stellar rotation axis, meaning we are probing along the equator with each transit. The stellar rotation period is $\sim$8x slower than the planet's orbital period. This means the planet makes multiple passes over the same starspots before they rotate out of view, as well as every 8'th transit approximately probing the same surface features. In this way we are able to measure the starspot properties on both shorter timescales from subsequent transits, and longer timescales due to the stellar rotation. 

\begin{figure}[!ht]
\makebox[\textwidth][c]{\includegraphics[width=7in]{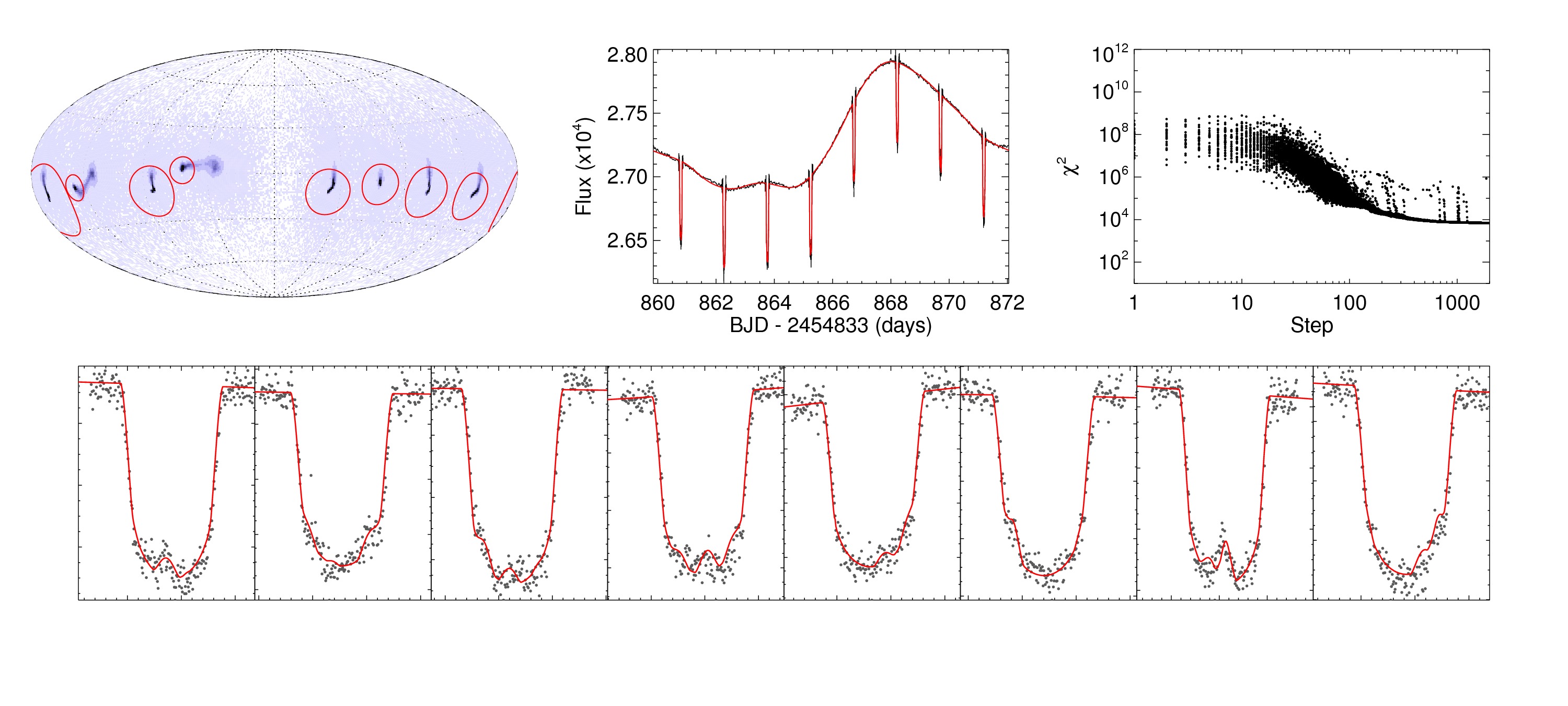}}
\caption{Results from our MCMC starspot modeling of a single 12.2 day time window for Kepler 17.  An eight-spot solution was used for this model. At top left is a contour map of the stellar surface, using a Hammer-Aitoff projection, showing the latitudes and longitudes of all eight spots over the entire MCMC run, with the best-fit spots scaled to their actual sizes (red circles). The light curve for the entire time window is shown (black dots, top center panel) with the best fit model overlaid (red line), with enlargements of all eight transits and the best fit model shown along the bottom. At top right is the $\chi^2$ for the entire MCMC run.}
\label{kep17}
\end{figure}

In Figure \ref{kep17} we show the results from modeling a single time window of Kepler 17, spanning one rotation period (12.2 days), and including eight transits. We have used the {\it Kepler} long cadence data for the out-of-transit portions of the light curve, and the short cadence data for the in-transit times. This example MCMC run used 1000 parameter space ``walkers'', running for 1000 steps, and was seeded with random spot positions and sizes. The best fit location for all eight starspots was close to the path of the planet (equator), creating bump features in every observed transit within this time window. These spots have their latitudes strongly constrained by the planetary transit path, thus breaking several fundamental degeneracies in the starspot modeling.

Note the small scale in-transit bumps and large scale out-of-transit modulations are both reproduced by our model. We are currently running these detailed starspot fitting models for every time window of Kepler 17 data, which spans 17 Quarters (four years) of {\it Kepler} data. To determine the number of starspots present within each time window we repeat our MCMC modeling run, each time increasing the numbers of spots. Thus we are able to recover the number of spots, each with the best fit latitude, longitude, and radius, as a function of time for transiting systems like this. Our preliminary results indicate that the starspot evolution timescales are considerably shorter than for GJ 1243, and that Kepler 17 has a more Solar-like rate of differential rotation.

\acknowledgments{
We acknowledge support for this work from NASA Kepler Cycle 2 GO grant NNX11AB71G, NASA Kepler Cycle 3 GO grant NNX12AC79G, and from NSF grant AST13-11678.
}

\end{document}